\documentclass[aps,prl,reprint,groupedaddress,showpacs,superscriptaddress,]{revtex4-1}
\usepackage{hyperref}
\usepackage{amssymb}
\usepackage{graphicx}
\usepackage{dcolumn}
\usepackage{bm}
\usepackage{amsmath}
\usepackage{fixmath}
\usepackage[usenames]{color}
\usepackage{float}
\usepackage{bbm}
\usepackage{epsfig}
\newcommand{\beq}{\begin{equation}}
\newcommand{\eeq}{\end{equation}}
\newcommand{\beqa}{\begin{eqnarray}}
\newcommand{\eeqa}{\end{eqnarray}}
\newcommand{\ba}{\begin{align}}
\newcommand{\ea}{\end{align}}

\newcommand{\iwn}{\text{i}\omega_n}
\newcommand{\BK}{{\bm{k}}}

\preprint{\fbox{\texttt{\tiny\jobname.tex}}}
\begin{document}
\title{Superconducting Phase and Pairing Fluctuations in the Half-Filled
Two-Dimensional Hubbard Model
}
\author{Michael Sentef}
\email[]{sentefmi@physik.uni-augsburg.de}
\affiliation{Theoretical Physics III, Center for Electronic Correlations and 
Magnetism, Institute of Physics, University of Augsburg, D-86135 Augsburg,
Germany}
\author{Philipp Werner}
\affiliation{Theoretische Physik, ETH Zurich, 8093 Z{\"u}rich, Switzerland}
\author{Emanuel Gull}
\affiliation{
Department of Physics, Columbia University, New York, New York 10027, USA}
\author{Arno P.\ Kampf}
\affiliation{Theoretical Physics III, Center for Electronic Correlations and 
Magnetism, Institute of Physics, University of Augsburg, D-86135 Augsburg,
Germany}
\date{\today}
%%%%%%%%%%%%%%%%%%%%%%%%%%%%%% 
%%%%%%%      ABSTRACT   %%%%%%%%%%%%%%
%%%%%%%%%%%%%%%%%%%%%%%%%%%%%%
\begin{abstract}
The two-dimensional Hubbard model exhibits superconductivity with $d$-wave symmetry even at half-filling in the presence of next-nearest neighbor hopping. Using plaquette cluster dynamical mean-field theory with a continuous-time quantum Monte Carlo impurity solver, we reveal the non-Fermi liquid character of the metallic phase in proximity to the superconducting state. Specifically, the low-frequency scattering rate for momenta near $(\pi,0)$ varies non-monotonously at low temperatures, and the dc conductivity is $T$-linear at elevated temperatures with an upturn upon cooling. Evidence is provided that pairing fluctuations dominate the normal-conducting state even considerably above the superconducting transition temperature.
\end{abstract}
\maketitle
%%%%%%%%%%%%%%%%%%%%%%%%%%%%%% 
%%%%%%%%%%% INTRO %%%%%%%%%%%%%%%
%%%%%%%%%%%%%%%%%%%%%%%%%%%%%%
The main postulate of Landau's Fermi-liquid (FL) theory \cite{AGD} is the existence of well-defined fermionic quasiparticles with a lifetime which diverges more strongly than $(E_{\bm{k}}-E_F)^{-1}$ when their energy $E_{\bm{k}}$ approaches the Fermi energy $E_F$. The search for instabilities of the FL towards alternative metallic states which are not adiabatically connected to the non-interacting Fermi gas has developed into one of the central topics in correlated electron physics. Perturbation theory of electron-electron interaction in dimensions higher than 1 reproduces the FL, implying that processes not contained therein are needed to destroy it \cite{Chubukov}. We will present evidence in this letter that in the two-dimensional (2D) Hubbard model (HM), a generic model for interacting electrons on a lattice, pairing fluctuations are such processes which eventually lead to a breakdown of FL theory even for moderate interaction strength in the proximity to a superconducting groundstate.
  
The 2D HM has been studied extensively in the intermediate-to-strong coupling regime believed to be relevant for the physics of doped Mott insulators and cuprate superconductors. In the cuprates, deviations from FL behavior in the metallic regime were observed both in quantum oscillation and angle-resolved photoemission experiments \cite{NormanNJP,Fournier}. Similarly, in the 2D HM the FL may become unstable towards antiferromagnetism (AF) \cite{Lin,Duffy} or $d$-wave superconductivity (dSC) \cite{DSCDoping1,DSCDoping2,DSCDoping3,DSCDoping4}. Non-Fermi liquid (NFL) signatures and anisotropic pseudogaps have been identified for electron densities below half-filling \cite{G,W,V,Ferr}. As we discuss in this letter, a state with properties reminiscent of the strongly correlated state below half-filling already appears in the seemingly simpler weakly coupled half-filled system with finite next-nearest neighbor hopping $t'$.

The possibility of a dSC instability at moderate coupling even at half-filling was previously discussed on the basis of renormalized mean-field \cite{Reiss} as well as cluster dynamical mean-field calculations \cite{Kancharla}. A weak-coupling renormalization group analysis confirmed the existence of a dSC instability with a critical temperature which vanishes exponentially towards zero interaction strength \cite{Raghu}. Functional renormalization group studies indicated the possibility of NFL normal state properties, in proximity to van Hove singularities \cite{fRGNFL1,fRGNFL2} and also in more extended parameter regimes \cite{Honerkamp02,Ossadnik}.

Here we study the half-filled 2D HM in the paramagnetic phase at finite but low temperatures. Using plaquette cluster dynamical mean-field theory (CDMFT), we confirm the existence of dSC at moderate coupling. We investigate the normal-state properties above the dSC critical temperature and reveal an unusual metallic state characterized by an increasing instead of decreasing single-particle scattering rate upon lowering the temperature $T$. Moreover, the dc resistivity is $T$-linear at elevated temperatures and shows an upturn at lower $T$. This evidence either implies that dSC emerges from an unusual metallic state with NFL character or that superconducting fluctuations themselves severely modify the metallic state above $T_c$.

%%%%%%%%%%%%%%%%%%%%%%%%%%%%%% 
%%%  MODEL AND METHODS   %%%%%%%%%%%%
%%%%%%%%%%%%%%%%%%%%%%%%%%%%%%
The 2D $U$-$t$-$t'$ Hubbard model reads
\begin{equation}
H = \sum_{\BK,\sigma} (\epsilon_{\BK}-\mu)
c^\dagger_{\BK,\sigma}c^{}_{\BK,\sigma}+U\sum_i n_{i,\uparrow}n_{i,\downarrow},
\label{H}
\end{equation} 
where $c^\dagger_{\BK,\sigma}$ ($c^{}_{\BK,\sigma}$) creates (annihilates) an electron in a Bloch state with momentum $\BK$ and dispersion
$
\epsilon_\BK=-2t\left(\cos k_x + \cos k_y \right)-4t'\cos k_x \cos k_y
$, $n_{i,\sigma}$ is the density operator for site $i$ and spin $\sigma$ $=$ $\uparrow,\downarrow$, $U$ $>$ 0 is the local Coulomb repulsion strength, and $\mu$ the chemical potential.
%Below we use the energy unit $t$ $=$ 1. 
%The model is solved using CDMFT for an $N_c$ $=$ $2 \times 2$ plaquette. 
The CDMFT self-consistency equations \cite{CDMFT1,CDMFT2} are
\beqa
\bm{G}(\iwn) &=&
\displaystyle
\sum_{\bm{\tilde{k}}}% \in \text{red.\ BZ}}
\left(
(\iwn+\mu)\bm{1}-\bm{\Sigma}(\iwn)
-
\bm{t}(\bm{\tilde{k}})
\right)^{-1},
\\
\bm{\mathcal{G}}_0^{-1}(\iwn) &=& 
\bm{G}^{-1}(\iwn)-\bm{\Sigma}(\iwn).
\eeqa
For the $N_c$ $=$ $2 \times 2$ plaquette, the hopping matrix $\bm{t}(\bm{\tilde{k}})$ is defined via its matrix elements $\bm{t}_{ij}(\bm{\tilde{k}})$ $=$ $N_c^{-1} \sum_{\bm{k}} e^{\text{i}(\bm{k}+\bm{\tilde{k}})\cdot(\bm{X}_i-\bm{X}_j)}$ $\epsilon_{\bm{k}+\bm{\tilde{k}}}$, where $\bm{X}_i$ and $\bm{X}_j$ are the position vectors of cluster sites $i$ and $j$, $\bm{\tilde{k}}$ is in the reduced Brillouin zone, and the cluster momentum takes the values $\bm{k}$ $=$ $(0,0)$, $(\pi,0)$, $(0,\pi)$, and $(\pi,\pi)$. All quantities, i.e. $\bm{t}$, the coarse-grained cluster Green function $\bm{G}$, the corresponding Weiss field $\bm{\mathcal{G}}_0$ and the cluster self-energy $\bm{\Sigma}$ are $N_c \times N_c$ matrices, and $\bm{1}$ is the unit matrix. In the following we consider only paramagnetic solutions and the spin index is therefore suppressed.
The self-consistency cycle is closed by solving the impurity problem, i.e. by calculating, for a given self-energy and the corresponding Weiss field, a new cluster Green function matrix $\bm{G}_{ij}(\tau) = -\langle\mathcal{T}_\tau c_i^{}(\tau)c_j^{\dagger}(0)\rangle_{S_{\text{eff}}}$, where $S_{\text{eff}}$ denotes the effective action of the auxiliary Anderson impurity model.

%%%%%%%%%%%%%%%%%%%%%%%%%%%%%
%%%%%% FIGURE 1 %%%%%%%%%%%%%%%%
%%%%%%%%%%%%%%%%%%%%%%%%%%%%%%
\begin{figure}[t]
\includegraphics[width=\hsize,angle=0]{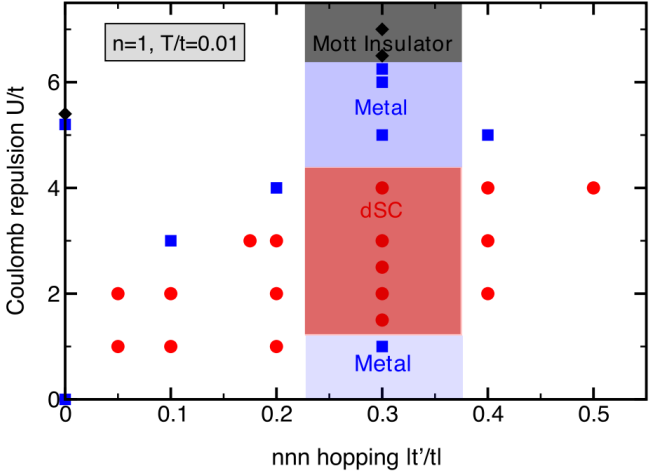}
\caption{
Overview of paramagnetic phases obtained for the half-filled 2D $U$-$t$-$t'$ Hubbard model within 2 $\times$ 2 CDMFT at $T/t$ $=$ 1/100. Data points show metallic (blue squares), dSC (red circles) and Mott insulator (black diamonds) states. For the characterization of phases see Fig.\ \ref{fig:overview}. The shaded areas are a guide to the eye. 
\label{fig:phase}}
\end{figure}
The CDMFT formalism is generalized to the superconducting state by introducing the spinors $\Psi_\BK = (c^{}_{\BK\uparrow}, c^{\dagger}_{-\BK\downarrow})$. The corresponding Nambu Green function matrix in momentum space reads
\beq
\hat{G}_{{\BK}}(\tau) = \langle \Psi^{}_\BK (\tau)  \Psi^{\dagger}_\BK \rangle= \left(\begin{array}{cc}
G_{\BK,\uparrow}(\tau) & F_{\BK}(\tau)
\\
F^{*}_{\BK}(\beta-\tau) & G_{-\BK,\downarrow}(\beta-\tau)
\end{array}\right).
\label{nambu}
\eeq
%The CDMFT formalism itself remains unchanged but the cluster Green function and self-energy become $2N_c \times 2N_c$ matrices, i.e. a 2 $\times$ 2 Nambu matrix for each cluster momentum. 
Within plaquette CDMFT the cluster Green function matrix is diagonal in cluster momentum space, and the relevant anomalous Green functions for $d$-wave order parameter symmetry are $F_{(\pi,0)}(\tau)$ $=$ $-F_{(0,\pi)}(\tau)$, while $F_{(0,0)}$ and $F_{(\pi,\pi)}$ both vanish. If a self-consistent dSC solution is obtained, the order parameter is 
$
\Delta_{\BK}$ $=$ $|F_\BK(\tau=0^+)|$ $=$ $|\langle c^{}_{\BK \uparrow}(0^+) c^{}_{-\BK \downarrow}\rangle|
$ for $\BK$ $=$ $(\pi,0)$.

%%%%%%%%%%%%%%%%%%%%%%%%%%%%%%
%%%%%%%%%%%%%%%%%%%%%%%%%%%%%% 
%%%%%%%%%%%  FIGURE 2   %%%%%%%%%%%%
%%%%%%%%%%%%%%%%%%%%%%%%%%%%%%
\begin{figure}[t]
\begin{center}
\includegraphics[width=\hsize]{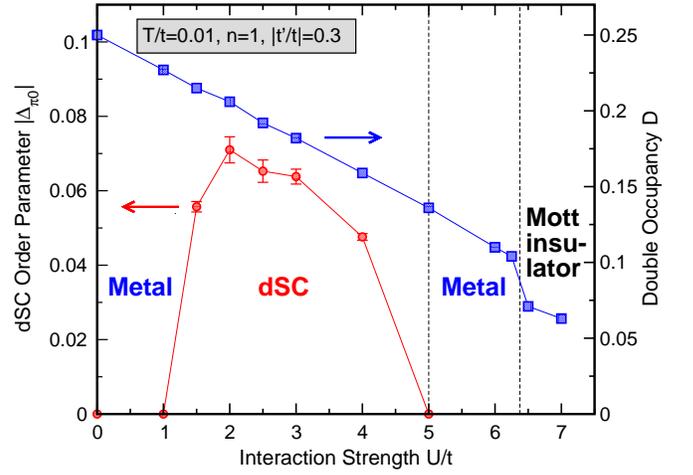} \\
\end{center}
\caption{
The superconducting order parameter $\Delta_{(\pi,0)}$ (left axis) and the double occupancy $D$ (right axis) as a function of $U/t$. For $U/t$ $\geq$ 6.5 only the Mott-insulating solution to the CDMFT equations is shown, but note that a metallic solution also exists in a hysteresis region of $U$ values.
\label{fig:overview}}
\end{figure}
%%%%%%%%%%%%%%%%%%%%%%%%%%%%%%
The impurity problem is solved by numerically exact continuous-time quantum Monte Carlo (QMC)  simulations based on the expansion of the effective action in the impurity-bath hybridization\cite{Werner06} generalized to superconducting states \cite{Haule07sc}. The hybridization expansion QMC allows to reach temperatures sufficiently low to detect dSC instabilities within plaquette CDMFT and to trace pairing fluctuations to temperatures well above $T_c$. 

We start with an overview of the plaquette CDMFT phases obtained for $T/t$ $=$ 0.01 in Fig.\ \ref{fig:phase}. Four distinct low-temperature phases are indicated: a metal at very weak coupling, a superconducting state with $d$-wave order parameter symmetry at moderate coupling, a metal with NFL properties, and a Mott insulator at strong coupling. Below we focus specifically on $|t'/t|$ $=$ 0.3 and analyze the distinctive character of the individual phases.

The average double occupancy $D$ $=$ $N_c^{-1}\sum_{i=1}^{N_c} \langle n_{i\uparrow} n_{i\downarrow} \rangle$ decreases with increasing $U$, until the discontinuous Mott transition\cite{Mottpapers1,Mottpapers2,Mottpapers3} is reached where a sharp drop of $D$ at $U_{c}$ between 6.25 and 6.5 occurs (see Fig.\ \ref{fig:overview}). A non-zero superconducting order parameter $|\Delta_{(\pi,0)}|$ is obtained in a converged dSC solution of the CDMFT equations in the range 1 $<$ $U/t$ $<$ 5. In the metallic phases the dSC order parameter vanishes. For even lower temperatures $T$ $<$ $t/100$ the dSC region in the phase diagram is expected to grow and most likely to extend to arbitrarily weak interaction strengths in the zero-temperature limit -- in conjunction with asymptotically exact weak-coupling renormalizaton group results \cite{Raghu}.

%%%%%%%%%%%%%%%%%%%%%%%%%%%%%% 
%%%%%%%%%%%  FIGURE 3   %%%%%%%%%%%%
%%%%%%%%%%%%%%%%%%%%%%%%%%%%%%
\begin{figure}[t]
\begin{center}
\includegraphics[width=\hsize]{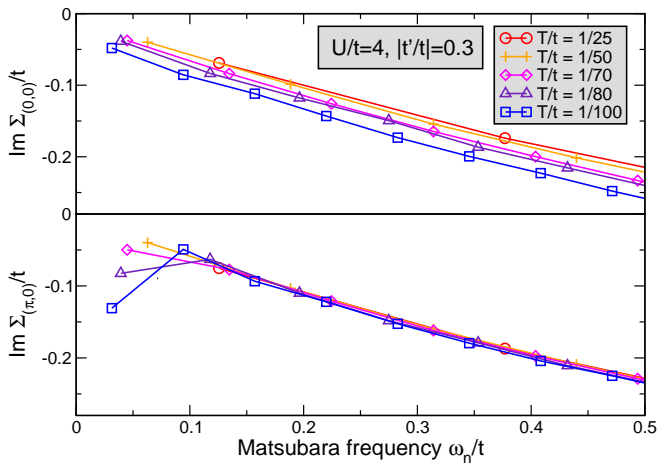} \\
\end{center}
\caption{
Imaginary part of the self-energy at $\BK$ $=$ $(0,0)$ and $(\pi,0)$ at the Matsubara frequencies $\iwn$ $=$ $\text{i}(2n+1)\pi T$ for $U/t$ $=$ 4, $|t'/t|$ $=$ 0.3, and different temperatures. Even though a dSC state is found for these parameters at $T/t$ $=$ $1/100$ we show here the self-energies for a solution of the CDMFT equations where only the diagonal Green functions are non-zero in Eq.\ (\ref{nambu}). 
\label{fig:sgk}}
\end{figure}
%%%%%%%%%%%%%%%%%%%%%%%%%%%%%
In Fig.\ \ref{fig:sgk} we show the imaginary part of the self-energy $\text{Im}\Sigma_{\BK}(\iwn)$ for the cluster momenta $\BK$ $=$ $(0,0)$ and $(\pi,0)$ for $U$ $=$ $4t$ in the metallic phase in proximity to superconductivity. At $\BK$ $=$ $(0,0)$, i.e. far away from the Fermi surface, the absolute value of $\text{Im} \Sigma$ grows upon lowering the temperature below $T/t$ $=$ $1/50$; dSC sets in below $T/t$ $\approx$ $1/90$. The same behavior is also observed for stronger coupling $U/t$ $=$ 5 and for $t'$ $=$ 0 at $U/t$ $=$ 4. 

The self-energy at $(\pi,0)$, the closest point to the Fermi surface resolved in plaquette CDMFT, instead is remarkably different. In a 2D Fermi liquid the generic low-frequency form of the self-energy\cite{2dfl} $\text{Im}\Sigma(\omega)$ $\propto$ $\omega^2 \ln \omega + T^2 \ln T$ implies that also $\text{Im}\Sigma(\iwn)$ has a negative slope for $\omega_n$ $\rightarrow$ $0$, with a zero-frequency offset which vanishes for $T$ $\rightarrow$ 0. In contrast, the self-energy at $\BK$ $=$ $(\pi,0)$, shown in the lower panel of Fig.\ \ref{fig:sgk}, develops a low-frequency hump feature with a change in slope upon lowering $T$, indicative of an increased scattering rate near the Fermi energy. This hump feature is distinct from Fermi liquid behavior; it emerges uniquely in the parameter regime in which dSC occurs. For example, if $t'$ $=$ 0 or $U/t$ $\geq$ 5 the dSC instability and the NFL hump feature in the self-energy at $(\pi,0)$ are both absent. Within our numerical approach it is, however, difficult to trace this low-energy NFL anomaly to even lower $T$ or smaller $U$ values. At $\BK$ $=$ $(\pi,\pi)$, Im $\Sigma(\iwn)$ is almost independent of temperature and vanishes linearly at low frequencies.

These features for the normal-state self-energy suggest unusual transport properties, too. We therefore calculate the dc resistivity by performing a convolution on the imaginary frequency axis without vertex corrections \cite{Pruschke}. The longitudinal dc conductivity is obtained approximately via
\beqa
\sigma_{\text{dc}} &\simeq& \frac{e^2}{T^2 \pi h} \Lambda_{xx}(\tau={1}/{2T})
\label{eq:sigma1}
\eeqa
from the long-wavelength limit of the current-current correlation function
\beqa
\Lambda_{xx}(\tau) &=& \frac{T^2}{N}\hspace{-1mm}
\sum_{nm\bm{k}} (\partial_{k_x}\epsilon_\BK)^2
e^{-\text{i}\omega_m\tau} G_{\bm{k}}(\text{i}\omega_{n+m})
G_{\bm{k}}(\iwn)
\label{eq:sigma2}
\eeqa
for bosonic Matsubara frequencies $\omega_m$ $=$ $2m\pi T$. $N$ is the number of $k$ points in the first Brillouin zone, $G_{\bm{k}}(\iwn)$ $=$ $(\iwn+\mu-\epsilon_{\bm{k}}-\Sigma_{\bm{k}}(\iwn))^{-1}$, and $\Sigma_{\bm{k}}$ $=$ $\frac{1}{4}\sum_{ij}e^{\text{i}\bm{k}\cdot(\bm{X}_{i}-\bm{X}_{j})}\Sigma_{ij}$ is an interpolated lattice self-energy obtained from the cluster impurity self-energy $\Sigma_{ij}$ \cite{CDMFT1}.
The basis for the approximate Eq.\ (\ref{eq:sigma1}) is the relation between the current-current correlation function at imaginary time $\tau$ and the imaginary part of the current-current correlation function at real frequency $\omega$,
\begin{equation}
\Lambda_{xx}(\tau)=\int_{-\infty}^{\infty}\frac{d\omega}{\pi}\frac{\omega e^{-\omega\tau}}{1-e^{-\omega/T}}\;\frac{{\text{Im}}\Lambda_{xx}(\omega)}{\omega}.
\label{kernel}
\end{equation}
Within CDMFT $\Lambda_{xx}(\tau)$ can be directly obtained from Eq.\ (\ref{eq:sigma2}), but to determine $\sigma_{\rm{dc}}$ it is necessary to compute $\lim_{\omega \rightarrow 0}$ ${\text{Im}}\Lambda_{xx}(\omega)/\omega$ \cite{Scalapino}. 
However, $\frac{\omega e^{-\omega/2T}}{T(1-e^{-\omega/T})}$ is an even function, strongly peaked at $\omega/T$ $=$ 0 and decays exponentially in $\omega/T$. Therefore, $T^{-1}\Lambda_{xx}(\tau={1}/{2T})$ $\approx$ $\lim_{\omega \rightarrow 0}$ ${\text{Im}}\Lambda_{xx}(\omega)/\omega$ asymptotically for $T$ $\rightarrow$ 0; hence Eq.\ (\ref{eq:sigma1}) follows.

%%%%%%%%%%%%%%%%%%%%%%%%%%%%%% 
%%%%%%%%%%%  FIGURE 4   %%%%%%%%%%%%
%%%%%%%%%%%%%%%%%%%%%%%%%%%%%%
\begin{figure}[t]
\begin{center}
\includegraphics[width=\hsize]{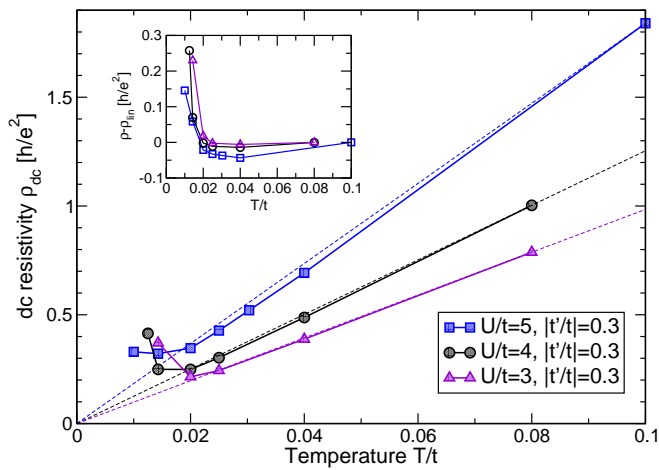} \\
\end{center}
\caption{
$T$-dependence of the dc resistivity $\rho_{\text{dc}}$ (without vertex corrections) for selected values of $U/t$. At high $T$ the resistivity is nearly linear in $T$, as indicated by the dashed lines ($\rho_{\text{lin}}$) which connect the origin and the data points at the respective highest temperatures. Deviations from this $T$-linear behavior set in below $T/t$ $\approx$ 0.02. Note that each curve stops at the smallest temperature at which the solution is not superconducting. Inset: Difference between $\rho_{\text{dc}}$ and $\rho_{\text{lin}}$.
\label{fig:rhot}}
\end{figure}
%%%%%%%%%%%%%%%%%%%%%%%%%%%%%%
The resulting $T$-dependent dc resistivity $\rho_{\text{dc}}$ $=$ $\sigma_{\text{dc}}^{-1}$ for selected values of $U/t$ is shown in Fig.\ \ref{fig:rhot}. Remarkably, the resistivity is $T$-linear at elevated temperatures. 
A deviation from the $T$-linear resistivity with an upturn at lower temperatures sets in at $T/t$ $\approx$ 0.02 for all $U$ values. Therefore, adding to the anomalous behavior of the antinodal $(\pi,0)$ self-energy, also the $T$-linear resistivity and the low-temperature upturn are different from Fermi liquid behavior. $\rho_{\text{dc}}(T)$ develops its upturn at approximately the same temperature at which the hump feature in the antinodal self-energy appears. 

%%%%%%%%%%%%%%%%%%%%%%%%%%%%%% 
%% CONCLUSIONS  %%%%%
%%%%%%%%%%%%%%%%%%%%%%%%%%%%%%
This hump bears resemblance with a similar feature observed in the self-energy on the real-frequency axis in the fRG results of Ref.\ \onlinecite{fRGNFL2}. The spike structure in Ref.\ \onlinecite{fRGNFL2} is especially pronounced near points on the Fermi surface which are connected by the umklapp scattering wave vector $\bm{Q}$ $=$ $(\pi,\pi)$. The origin of the spike feature in the fRG study is therefore attributed to enhanced umklapp scattering due to the emerging antiferromagnetic spin fluctuations. Similarly, also the NFL hump feature in our study suggests an explanation in terms of an emergent scattering mechanism. Since the $(\pi,0)$ self-energy develops its NFL structure uniquely for parameter values close to the dSC state, it is suggestive that they are caused by superconducting fluctuations, similarly as in the attractive Hubbard model in proximity to the superconducting state \cite{attractive}. 
Additional insight regarding the importance of superconducting fluctuations above $T_c$ may be gained from the probabilities of relevant plaquette eigenstates in the thermal ensemble \cite{GullEPL}. We have verified that a spin singlet two-hole state with zero total momentum, i.e. a ``Cooper pair state'' with respect to the half-filled plaquette, is represented with enhanced probability not only in the dSC state at low temperature, but already significantly above $T_c$. 

Despite the limitations imposed by neglecting current-vertex corrections and by the restricted momentum-space resolution, 
the $T$-linear behavior of the dc resistivity at elevated temperatures and 
its upturn at low temperatures mark nevertheless clear deviations from
Fermi liquid physics and bear resemblance to resistivity data in
cuprates \cite{upturn} and in other families of high-temperature superconductors as well \cite{transport}. Specifically superconducting phase fluctuations were recently verified in thermodynamic measurements well above $T_c$ and discussed in connection with the unusual properties of the metallic phase in the cuprates \cite{Li}. The resistivity upturn we find for the half-filled Hubbard model may therefore be tied to the superconducting fluctuations in the vicinity of the dSC phase. We note that the resistivity upturn occurs also for $U/t$ $=$ 5, but at a somewhat lower temperature than for the smaller $U/t$ values. This is in
accordance with Fig.\ \ref{fig:overview}, which shows that if superconductivity occurs for
$U/t$ $=$ 5, the critical temperature is smaller than $T/t$ $=$ 0.01.

Since we focused here on finite-temperature results for the 2D model, \emph{long-range} AF order was naturally not considered, while dSC can emerge via a Kosterlitz-Thouless type transition \cite{KT}. At $T$ $=$ 0, AF is strongest for $t'$ $=$ 0 with the lower critical coupling strength $U_c^{\text{AF}}$ $=$ 0 \cite{Lin,Duffy}. For $|t'|$ $>$ 0, long-range AF order is only obtained beyond a critical coupling $U_c^{\text{AF}}(t')$ $>$ 0 \cite{Duffy}. The RG results in Ref.\ \onlinecite{Raghu} imply the existence of a finite dSC range for $|t'|$ $>$ 0 and weak to moderate $U$ at $T$ $=$ 0. At $U_c^{\text{AF}}$ we expect a transition to a ground state with AF order, possibly in coexistence with dSC \cite{Kancharla,Reiss,Capone}.

Our findings show that even for moderate correlations the physics of the half-filled 2D Hubbard model with $t'$ $\neq$ 0 already contains the superconducting and the unusual metallic NFL phases encountered in previous studies of the doped system at stronger coupling. It seems likely that the strong-coupling dSC and NFL phases at finite hole doping are continuously connected to their counterparts at weak coupling and half-filling for finite $t'$ -- a hypothesis to be elaborated on in future work. Finally we note that our findings of NFL signatures also bear 
resemblance to the single-site DMFT results for the normal state of the 
attractive Hubbard model close above the transition to s-wave 
superconductivity \cite{attractive2}.

%%%%%%%%%%%%%%%%%%%%%%%%%%%%%% 
%%%%%%%%%%%% ACKNOWLEDGMENTS  %%%%%
%%%%%%%%%%%%%%%%%%%%%%%%%%%%%%
We thank Die\-ter Voll\-hardt, Pra\-buddha Cha\-kra\-bor\-ty, Michael Sekania, Walter Metzner, Carsten Honerkamp, and Maurice Rice for discussions. This work is supported by the DFG through TRR 80. M.S. acknowledges support by the Studienstiftung des Deutschen Volkes. E.G. acknowledges support by NSF-DMR-1006282 and P.W. by SNF grant PP001-118866. Computer simulations were performed on HLRB II at LRZ Garching using the ALPS libraries \cite{ALPS}.

%%%%%%%%%%%%%%%%%%%%%%%%%%%%%% 
%%% %%%%%%%%%%%% BIB  %%%%%%%%%%%%
%%%%%%%%%%%%%%%%%%%%%%%%%%%%%%

\end{document}